\documentclass[a4paper]{article}

\pdfoutput=1
\usepackage{INTERSPEECH2021}
\usepackage{url}
\usepackage{hyperref}
\usepackage{multirow}
\usepackage{xparse}
\NewDocumentCommand\alembic{}{\includegraphics[height=1em]{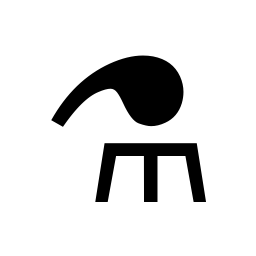}}

\title{Keyword Transformer: A Self-Attention Model for Keyword Spotting}
\name{Axel Berg$^{1,2, *}$, Mark O'Connor$^{1,*}$ \thanks{*Equal contribution.}, Miguel Tairum Cruz$^{1}$}

\address{
  $^1$Arm ML Research Lab, $^2$Lund University}
\email{\{axel.berg, mark.oconnor, miguel.tairum-cruz\}@arm.com}

\begin{document}

\maketitle
\begin{abstract}
The Transformer architecture has been successful across many domains, including natural language processing, computer vision and speech recognition. In keyword spotting, self-attention has primarily been used on top of convolutional or recurrent encoders. We investigate a range of ways to adapt the Transformer architecture to keyword spotting and introduce the Keyword Transformer (KWT), a fully self-attentional architecture that exceeds state-of-the-art performance across multiple tasks without any pre-training or additional data. Surprisingly, this simple architecture outperforms more complex models that mix convolutional, recurrent and attentive layers. KWT can be used as a drop-in replacement for these models, setting two new benchmark records on the Google Speech Commands dataset with 98.6\% and 97.7\% accuracy on the 12 and 35-command tasks respectively.\footnote{Code is available at \href{https://github.com/ARM-software/keyword-transformer}{https://github.com/ARM-software/keyword-transformer}.}
\end{abstract}

\section{Introduction}

Recent works in machine learning show that the Transformer architecture, first introduced by Vaswani et al.\ \cite{vaswani2017attention}, is competitive not only in language processing, but also in e.g.\ image classification, \cite{dosovitskiy2020image, touvron2020training, yuan2021tokens}, image colorization \cite{kumar2021colorization}, object detection \cite{carion2020end}, automatic speech recognition \cite{gulati2020conformer, chenwu2020streaming, liu2020tera}, video classification \cite{neimark2021video} and multi-agent spatiotemporal modeling \cite{alcorn2021baller2vec}. This can be seen in the light of a broader trend, where a single neural network architecture generalizes across many domains of data and tasks.

Attention mechanisms have also been explored for keyword spotting \cite{de2018neural, rybakov2020streaming}, but only as an extension to other architectures, such as convolutional or recurrent neural networks.

Inspired by the strength of the simple Vision Transformer (ViT) model \cite{dosovitskiy2020image} in computer vision and by the techniques that improves its data-efficiency \cite{touvron2020training}, we propose an adaptation of this architecture for keyword spotting and find that it matches or outperforms existing models on the much smaller Google Speech Commands dataset \cite{warden2018speech} without additional data.

\begin{figure}[t]
  \centering
  \includegraphics[width=\linewidth]{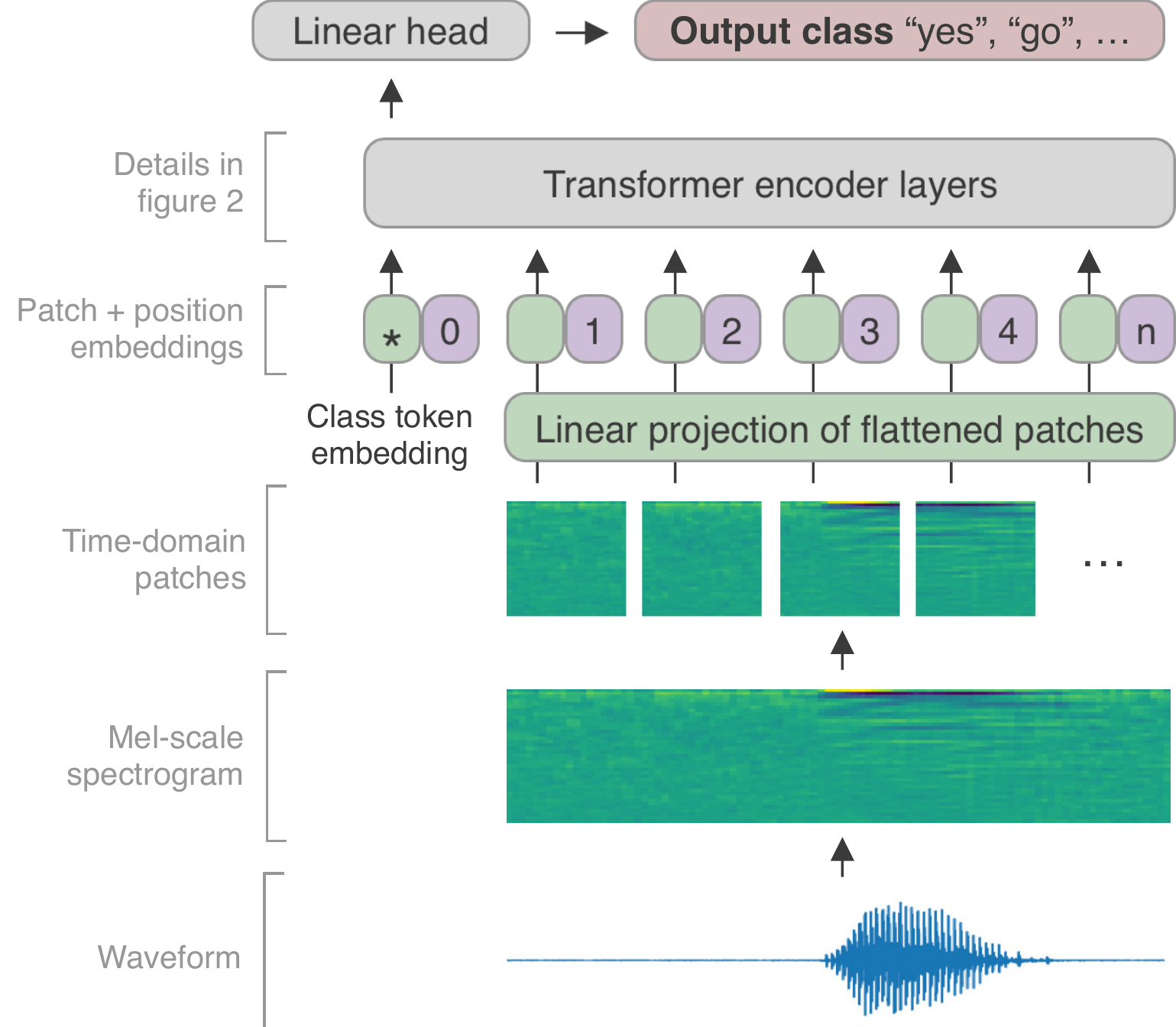}
  \caption{The Keyword Transformer architecture. Audio is preprocessed into a mel-scale spectrogram, which is partitioned into non-overlapping patches in the time domain. Together with a learned class token, these form the input tokens for a multi-layer Transformer encoder. As with ViT \cite{dosovitskiy2020image}, a learned position embedding is added to each token. The output of the class token is passed through a linear head and used to make the final class prediction.}
  \label{fig:overview}
\end{figure}

We summarize our main contributions as follows:
\begin{enumerate}
\item An investigation into the application of the Transformer architecture to keyword spotting, finding that applying self-attention is more effective in the time domain than in the frequency domain.
\item We introduce the Keyword Transformer, as illustrated in Figure \ref{fig:overview}, a fully self-attentional architecture inspired by ViT \cite{dosovitskiy2020image} that can be used as a drop-in replacement for existing keyword spotting models and visualize the effect of the learned attention masks and positional embeddings.
\item An evaluation of this model across several tasks using the Google Speech Commands dataset with comparisons to state-of-the-art convolutional, recurrent and attention-based models.
\item An analysis of model latency on a mobile phone, showing that the Keyword Transformer is competitive in edge use cases.
\end{enumerate}

\section{Related Work}

\begin{figure}[t]
  \centering
  \includegraphics[width=\linewidth]{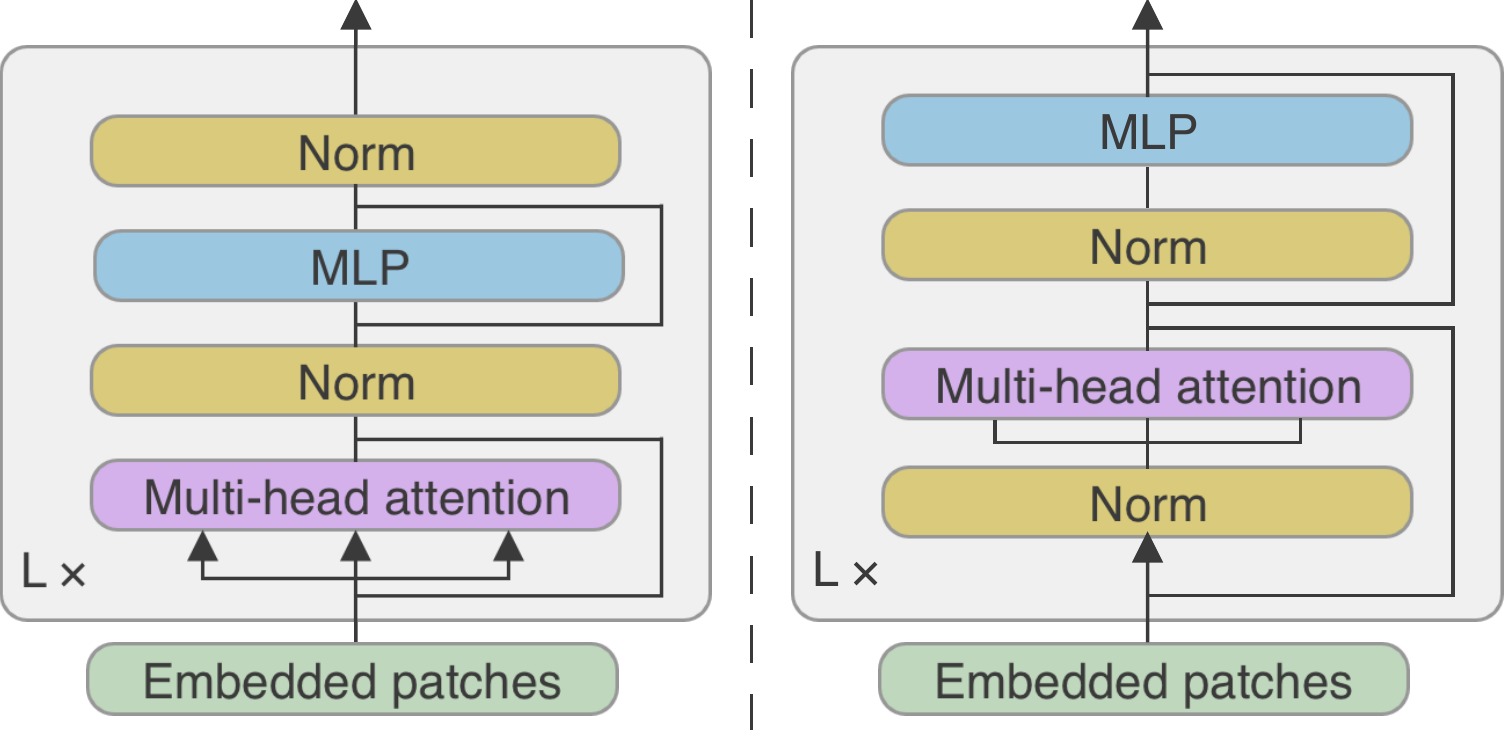}
  \caption{The PostNorm (left) and PreNorm (right) Transformer encoder architectures. KWT uses a PostNorm encoder.}
  \label{fig:norm}
\end{figure}

\subsection{Keyword Spotting}

Keyword spotting is used to detect specific words from a stream of audio, typically in a low-power always-on setting such as smart speakers and mobile phones. To achieve this, audio is processed locally on the device. In addition to detecting target words, classifiers may also distinguish between ``silence'' and ``unknown'' for words or sounds that are not in the target list.

In recent years, machine learning techniques, such as deep (DNN), convolutional (CNN), recurrent (RNN) and Hybrid-Tree \cite{MLSYS2019_a97da629} neural networks, have proven to be useful for keyword spotting. These networks are typically used with a pre-processing pipeline that extracts the mel-frequency cepstrum coefficients (MFCC) \cite{davis1980comparison}. Zhang et al.\ \cite{zhang2017hello} investigated several small-scale network architectures and identified depthwise-separable CNN (DS-CNN) as providing the best classification/accuracy tradeoff for memory footprint and computational resources. Other works have improved upon this result using synthesized data \cite{lin2020training}, temporal convolutions \cite{choi2019temporal, majumdar2020matchboxnet}, and self-attention \cite{de2018neural}. Recently Rybakov et al.\ \cite{rybakov2020streaming}  achieved a new state of the art result on Google Speech Commands using MHAtt-RNN, a non-streaming CNN, RNN and multi-headed (MH) self-attention model.

\subsection{Self-Attention and the Vision Transformer}

Dosovitskiy et al.\ introduced the Vision Transformer (ViT) \cite{dosovitskiy2020image} and showed that Transformers can learn high-level image features by computing self-attention between different image patches. This simple approach outperformed CNNs but required pre-training on large datasets. Touvroun et al.\ \cite{touvron2020training} improved data efficiency using strong augmentation, careful hyperparameter tuning and token-based distillation.

While Transformers have been explored for wake word detection \cite{wang2021wake} and voice triggering \cite{adya2020hybrid}, to the best of our knowledge fully-attentional models based on the Transformer architecture have not been investigated for keyword spotting. Our approach is inspired by ViT, in the sense that we use patches of the audio spectrogram as input and closely follows \cite{touvron2020training} to understand how generally this technique applies to new domains. We restrict ourselves to a non-streaming setting in this work, noting that others have previously investigated extensions of Transformers to a streaming setting \cite{wang2021wake, chenwu2020streaming}.

\section{The Keyword Transformer}

\subsection{Model Architecture}

Let $X \in \mathbb{R}^{T \times F}$ denote the output of the MFCC spectrogram, with time windows $t = 1,..., T$ and frequencies $f = 1, ..., F$. The spectrogram is first mapped to a higher dimension $d$, using a linear projection matrix $W_0 \in \mathbb{R}^{F \times d}$ in the frequency domain. In order to learn a global feature that represents the whole spectrogram, a learnable class embedding $X_{\text{class}} \in \mathbb{R}^{1 \times d}$ is concatenated with the input in the time-domain. Then a learnable positional embedding matrix $X_{\text{pos}} \in \mathbb{R}^{(T+1) \times d}$ is added, such that the input representation fed into the Transformer encoder is given by
\begin{equation}
X_0 = [ X_{\text{class}};XW_0 ] + X_{\text{pos}}
\end{equation}
The projected frequency-domain features are then fed into a sequential Transformer encoder consisting of $L$ multi-head attention (MSA) and multi-layer perceptron (MLP) blocks. In the $l$:th Transformer block, queries, keys and values are calculated as $Q = X_lW_Q$, $K = X_lW_K$ and $V = X_lW_V$ respectively, where $W_Q, W_K, W_V \in \mathbb{R}^{d \times d_h}$ and $d_h$ is the dimensionality of each attention-head. The self attention (SA) is calculated as
\begin{equation}
\text{SA}(X_l) = \text{Softmax}\Big(\frac{QK^T}{\sqrt{d_h}}\Big)V
\end{equation}
The MSA operation is obtained by linearly projecting the concatenated output, using another matrix $W_P \in \mathbb{R}^{k d_h \times d}$, from the $k$ attention heads.
\begin{equation}
\text{MSA}(X_l) = [\text{SA}_1(X_l); \text{SA}_2(X_l); ...; \text{SA}_k(X_l)] W_P
\end{equation}
In our default setting, we use the PostNorm \cite{vaswani2017attention} Transformer architecture as shown in Figure \ref{fig:norm}, where the Layernorm (LN) \cite{ba2016layer} is applied after the MSA and MLP blocks, in contrast to the PreNorm \cite{he2016deep} variant, where LN is applied first. This decision is discussed further in the ablation section. As is typical for Transformers, we use GELU \cite{hendrycks2016gaussian} activations in all MLP blocks.

In summary, the output of the $l$:th Transformer block is given by
\begin{align}
\tilde{X}_l &= \text{LN}(\text{MSA}(X_{l-1}) + X_{l-1}), \quad &l = 1,...,L \\
X_l &= \text{LN}(\text{MLP}(\tilde{X}_l) + \tilde{X}_l), \quad &l = 1, ..., L
\end{align}
At the output layer, the class embedding is fed into a linear classifier. Our approach treats time windows in a manner analogous to the handling of image patches in ViT. Whereas in ViT, the self-attention is computed over image patches, the attention mechanism here takes place in the time-domain, such that different time windows will attend to each other in order to form a global representation in the class embedding.

The model size can be adjusted by tuning the parameters of the Transformer. Following \cite{touvron2020training}, we fix the number of sequential Transformer encoder blocks to 12, and let $d/k = 64$, where $d$ is the embedding dimension and $k$ is the number of attention heads. By varying the number of heads as $k = 1, 2, 3$, we end up with three different models as shown in Table \ref{tab:params}.

\begin{table}[t]
	\caption{Model parameters for the KWT architecture.}
	\label{tab:params}
	\centering
	\begin{tabular}{llllll}
	\toprule
	\textbf{Model} & dim & mlp-dim & heads & layers &  \# parameters \\
	\midrule
	KWT-1 & 64 & 256 & 1 & 12 & 607K \\
	KWT-2 & 128 & 512 & 2 & 12 & 2,394K \\
	KWT-3 & 192 & 768 & 3 & 12 & 5,361K \\
	\bottomrule
	\end{tabular}
\end{table}

\subsection{Knowledge Distillation}

As introduced by Hinton et al.\ \cite{hinton2015distill}, knowledge distillation uses a pre-trained teacher's predictions to provide an auxiliary loss to the student model being trained. Touvron et al.\ \cite{touvron2020training}, introduced a distillation token, finding this benefits Transformers in the small data regime. This method adds a learned distillation token to the input. At the output layer this distillation token is fed into a linear classifier and trained using hard (one-hot) labels predicted by the teacher.

Let $Z_\mathrm{sc}$ be the logits of the student class token, $Z_\mathrm{sd}$ be the logits of the student distillation token and $Z_\mathrm{t}$ be the logits of the teacher model. The overall loss becomes
\begin{align}
    \mathcal{L} = \frac{1}{2}\mathcal{L}_\mathrm{CE}(\psi(Z_\mathrm{sc}),y) + \frac{1}{2}\mathcal{L}_\mathrm{CE}(\psi(Z_\mathrm{sd}),y_\mathrm{t}),
\end{align}
where $y_\mathrm{t}=\mathrm{argmax}(Z_\mathrm{t})$ are the hard decision of the teacher, $y$ are the ground-truth labels, $\psi$ is the softmax function and $\mathcal{L}_\mathrm{CE}$ is the cross-entropy loss. At inference time the class and distillation token predictions are averaged to produce a single prediction. Note that unlike Noisy Student \cite{xie2020student}, the teacher receives the same augmentation of the input as the student, effectively correcting labels made invalid by very strong augmentation. In all experiments, we use MHAtt-RNN as a teacher and denote distillation models with KWT\alembic.

\section{Experiments}

\begin{table}[t]
\scriptsize
\caption{Hyperparameters used in all experiments.}
\label{tab:hyperparams}
\centering
\begin{tabular}{ll}
\toprule
\multicolumn{2}{c}{\textbf{Training}} \\
\midrule
Training steps & 23,000 \\
Batch size & 512 \\
Optimizer & AdamW \\
Learning rate & 0.001 \\
Schedule & Cosine \\
Warmup epochs & 10 \\
\midrule
\multicolumn{2}{c}{\textbf{Regularization}} \\
\midrule
Weight decay & 0.1 \\
Label smoothing & 0.1 \\
Dropout & 0 \\ 
\bottomrule
\end{tabular}
\qquad
\begin{tabular}{ll}
\toprule
\multicolumn{2}{c}{\textbf{Pre-processing}} \\
\midrule
Time window length & 30 ms \\
Time window stride & 10 ms \\
\#DCT Features & 40 \\
\midrule
\multicolumn{2}{c}{\textbf{Data augmentation}} \\
\midrule
Time shift [ms] & [-100, 100] \\
Resampling & [0.85, 1.15] \\
Background vol. & 0.1 \\
\#Time masks & 2 \\
Time mask size & [0,25] \\
\#Frequency masks & 2 \\
Frequency mask size & [0,7] \\
\bottomrule
\end{tabular}
\end{table}

\subsection{Keyword Spotting on Google Speech Commands}

We provide experimental results on the Google Speech Commands dataset V1 and V2 \cite{warden2018speech}. Both datasets consist of 1 second long audio snippets, sampled at 16 kHz, containing utterances of short keywords recorded in natural environments. V1 of the dataset contains 65,000 snippets of 30 different words, whereas V2 contains 105,000 snippets of 35 different words. The 12-label classification task uses 10 words:  ”up”, ”down”, ”left”, ”right”, ”yes”, ”no”, ”on”, ”off”, ”go”, and ”stop”, in addition to "silence" and "unknown", where instances of the latter is taken from the remaining words in the dataset, whereas the 35-label task uses all available words. We use the same 80:10:10 train/validation/test split as \cite{warden2018speech, zhang2017hello, rybakov2020streaming} for side-by-side comparisons. We adhere as closely as possible to the evaluation criteria of \cite{rybakov2020streaming}, and for each experiment, we train the model three times with different random initializations. 

As our intention is to explore the extent to which results using Transformers from other domains transfer to keyword spotting, we follow the choices and hyperparameters from \cite{touvron2020training} as closely as possible, with the notable exception that we found increasing weight decay from 0.05 to 0.1 to be important. Furthermore, we use the same data pre-processing and augmentation policy as in \cite{rybakov2020streaming}, which consists of random time shifts, resampling, background noise, as well as augmenting the MFCC features using SpecAugment \cite{park2019specaugment}. We train our models over the same number of total input examples as MHAtt-RNN (12M) to allow a fair comparison. For clarity, the hyperparameters used in all experiments are reported in Table \ref{tab:hyperparams}.

The results are shown in Table \ref{tab:results}, where for our own results, we report a 95\% confidence interval for the mean accuracy over all three model evaluations. Our best models match or surpass the previous state-of-the-art accuracies, with significant improvements on both the 12-label and 35-label V2-datasets. In general, Transformers tend to benefit more from large amounts of data, which could explain why KWT does not outperform MHAtt-RNN on the smaller V1-dataset. Nevertheless, we also note that knowledge distillation is effective in improving the accuracy of KWT in most scenarios.

\begin{table}[t]
\footnotesize
  \caption{Accuracy on Speech Commands V1 \cite{speechv1} and V2 \cite{speechv2}.}
  \label{tab:results}
  \centering
  \resizebox{\columnwidth}{!}{\begin{tabular}{lllll}
    \toprule
    \textbf{Model}  & \textbf{V1-12} & \textbf{V2-12} & \textbf{V2-35} \\
    \midrule
    DS-CNN \cite{zhang2017hello} & 95.4 & & \\
    TC-ResNet \cite{choi2019temporal} & 96.6 & & \\
    Att-RNN \cite{de2018neural} & 95.6 & 96.9 & 93.9 \\
    MatchBoxNet \cite{majumdar2020matchboxnet} & 97.48 $\pm 0.11$ & 97.6 & \\
    Embed + Head \cite{lin2020training} & & 97.7 & \\
    MHAtt-RNN \cite{rybakov2020streaming}      & 97.2 & 98.0 &  \\
    Res15 \cite{vygon2021learning} & & 98.0 & 96.4 \\
    \midrule
    MHAtt-RNN (Ours)		& \textbf{97.50} $\pm 0.29$ & 98.36 $\pm 0.13$  & 97.27 $\pm 0.02$ \\
	KWT-3 (Ours)		& 97.24$\pm 0.24$ & \textbf{98.54} $\pm 0.17$ & 97.51 $\pm 0.14$ \\
	KWT-2 (Ours)		& 97.36 $\pm 0.20$ & 98.21 $\pm 0.06$  & 97.53 $\pm 0.07$  \\
	KWT-1 (Ours)		& 97.05 $\pm 0.23$ & 97.72 $\pm 0.01$ & 96.85 $\pm 0.07$ \\
    \midrule
	KWT-3\alembic (Ours)	& \textbf{97.49} $\pm 0.15 $ & \textbf{98.56} $\pm 0.07$ & 97.69 $\pm 0.09$ \\
	KWT-2\alembic (Ours)	& 97.27 $\pm 0.08$ & 98.43 $\pm 0.08$ & \textbf{97.74 $\pm 0.03$} \\
	KWT-1\alembic (Ours)		& 97.26 $\pm 0.18$ & 98.08 $\pm 0.10$ & 96.95 $\pm 0.14$ \\
	\bottomrule
  \end{tabular}}
\end{table}

\subsection{Ablation Studies}

\begin{figure}[b]
  \centering
  \includegraphics[width=\linewidth]{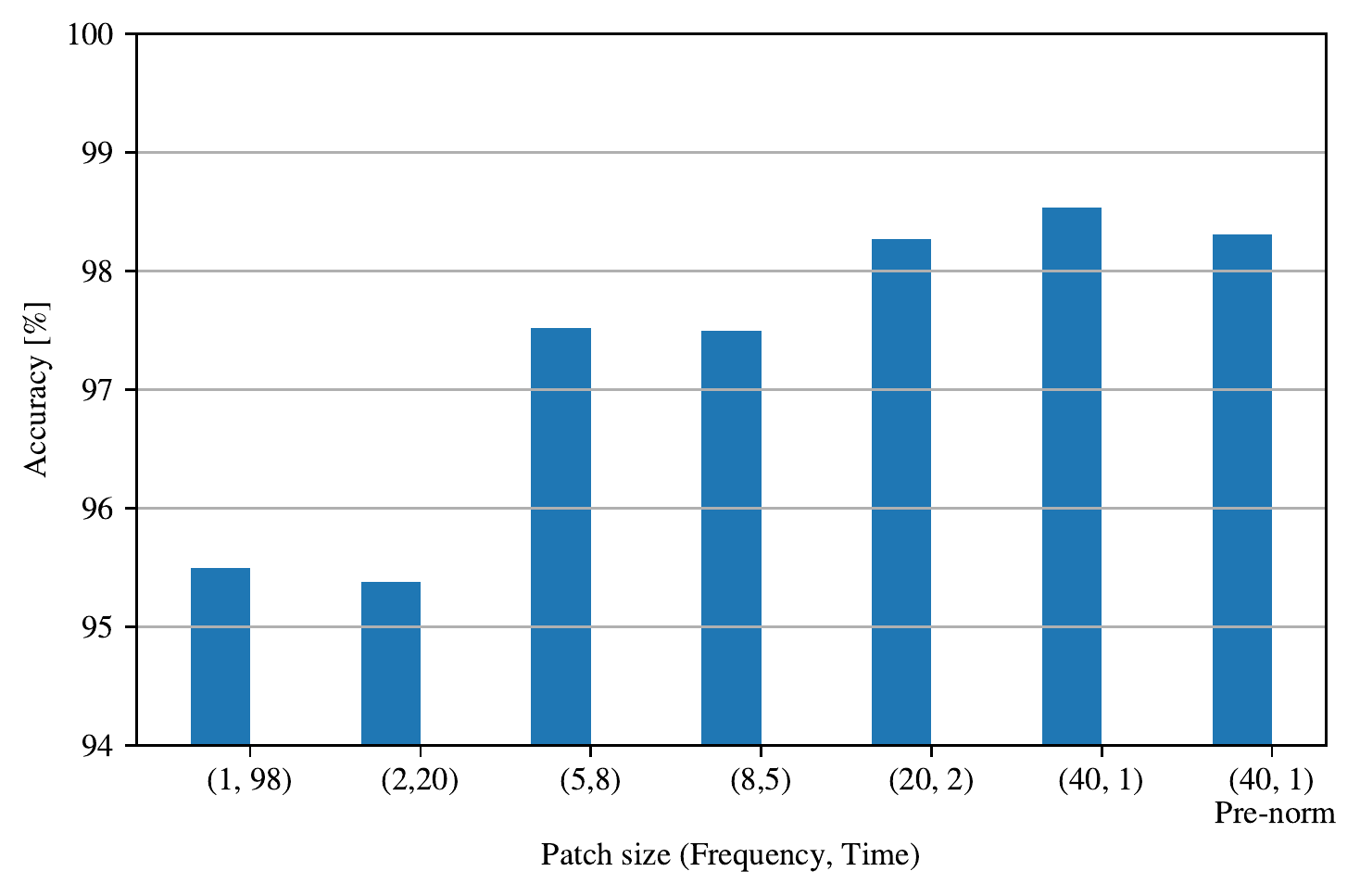}
  \caption{Accuracy on Speech Commands V2-12 using KWT-3 with different patch sizes.}
  \label{fig:ablation}
\end{figure}

We investigate different approaches to self-attention by varying the shapes of the MFCC spectrogram patches that are fed into the Transformer. Using our default hyperparameters, the spectrogram consists of 98 time windows, containing 40 mel-scale frequencies. Our baseline uses time-domain attention, but we also investigate frequency-domain attention and intermediate steps where rectangular patches are used. We find time-domain attention to perform best, as shown in Figure \ref{fig:ablation}. This is in agreement with previous findings that temporal convolutions work well for keyword spotting \cite{choi2019temporal}, since the first projection layer of our model can be interpreted as a temporal convolution with kernel size (40, 1) and stride 1 in the time-domain.

We also investigate the use of PreNorm and PostNorm and found that the latter improves performance for keyword spotting in our experiments. This is contrary to previous findings on other tasks \cite{nguyen2019Transformers}, where PreNorm has been shown to yield better results and we encourage further work to explore the role of normalization in Transformers across different domains.

\subsection{Attention Visualization}
\begin{figure}[t]
  \centering
  \includegraphics[width=\linewidth]{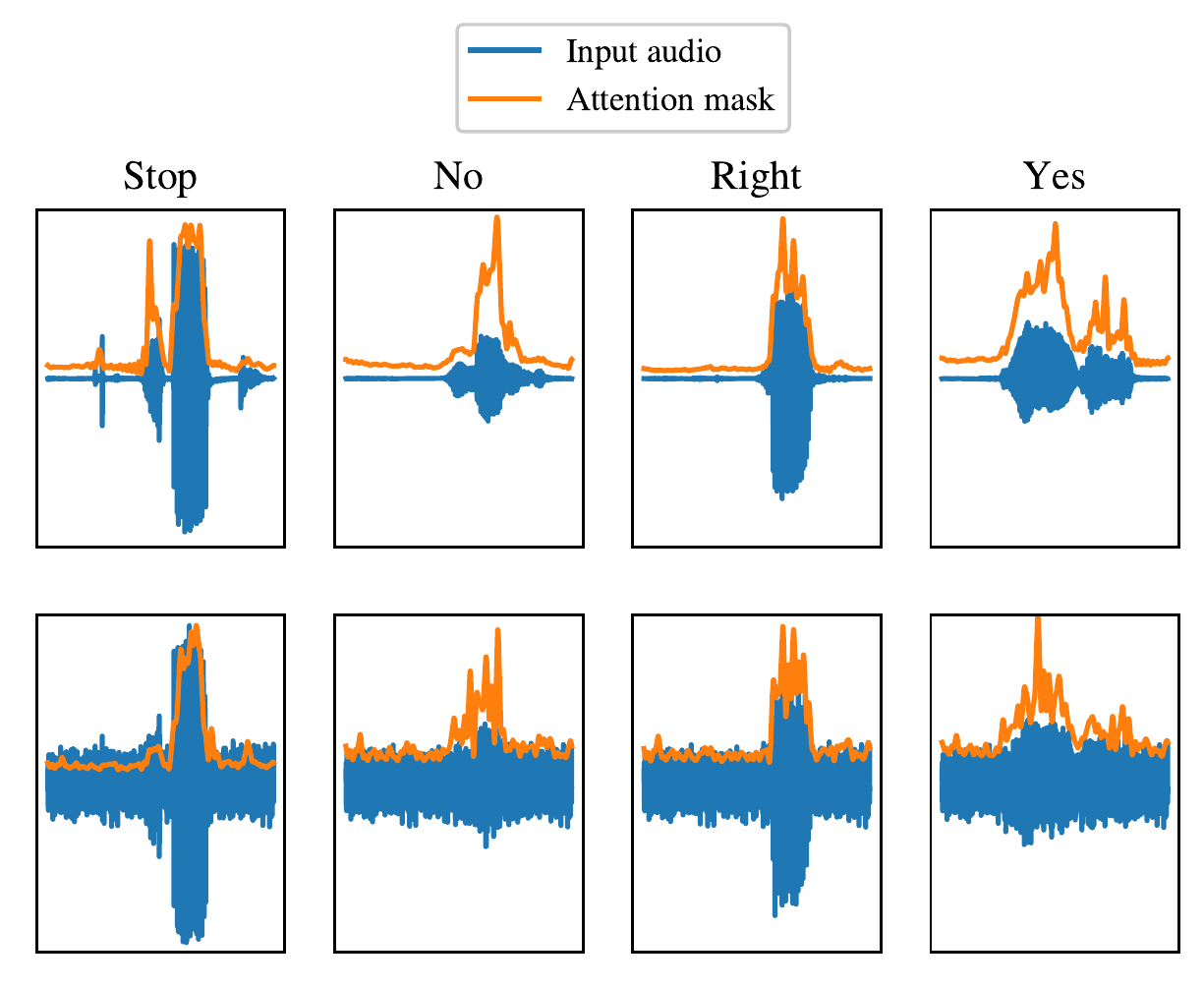}
  \caption{The learned attention mask, propagated from the input to the class token, overlaid on four different audio snippets, without (top) and with (bottom) background noise.}
  \label{fig:mask}
\end{figure}

In order to examine which parts of the audio signal the model attends to, we propagate the attention weights of each Transformer layer from the input to the class token by averaging the attention weights over all heads. This produces a set of attention weights for each time window of the input signal. Figure \ref{fig:mask} shows the attention mask overlayed on the waveform of four different utterances. It can be seen that the model is able to pay attention to the important parts of the input while effectively suppressing  background noise. 

\begin{figure}[b]
  \centering
  \includegraphics[width=\linewidth]{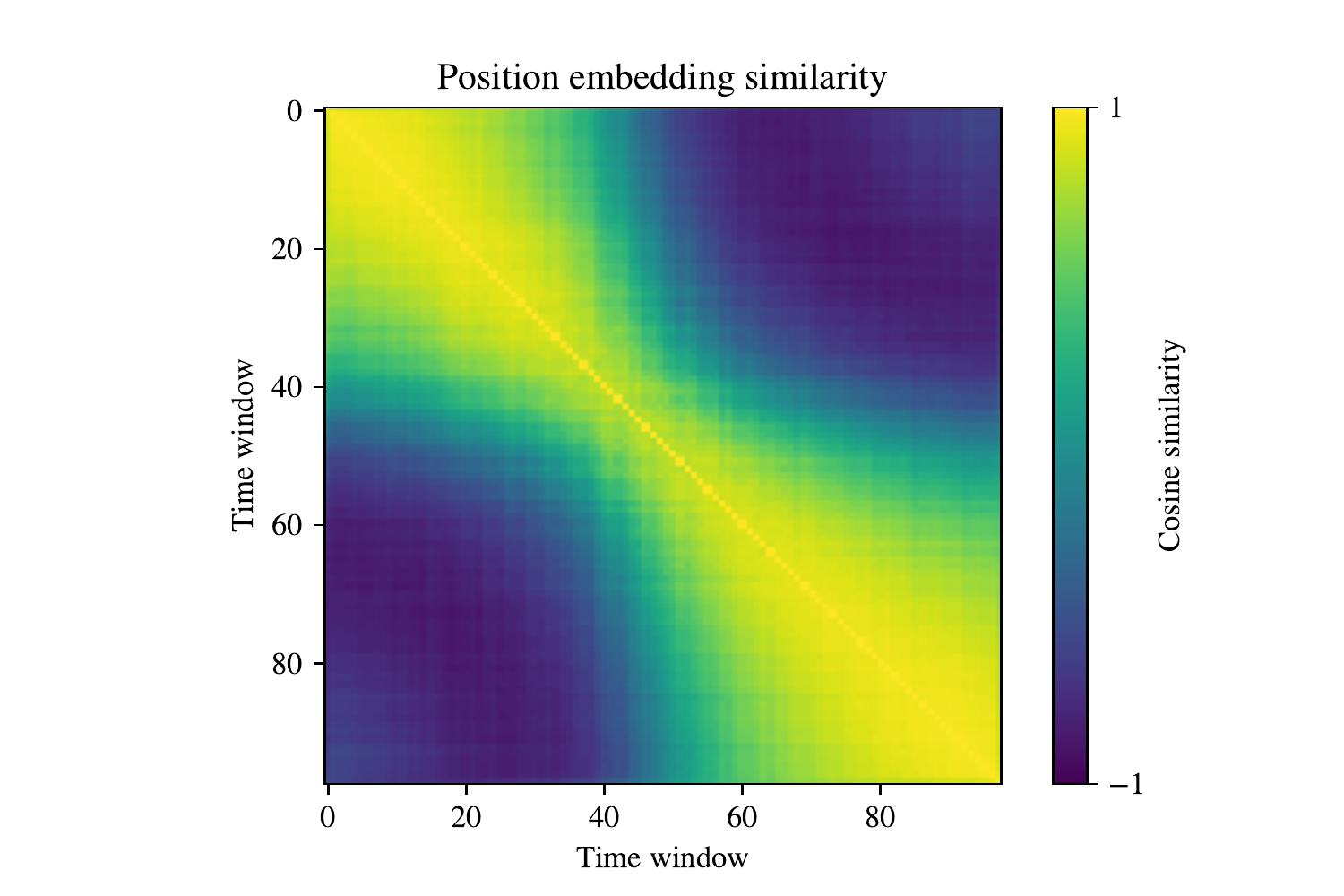}
  \caption{Cosine similarities of the learned position embeddings of KWT.}
  \label{fig:pos_similar}
\end{figure}

We also study the position embeddings of the final model by analyzing their cosine similarity, as shown in Figure \ref{fig:pos_similar}. Nearby position embeddings exhibit a high degree of similarity and distant embeddings are almost orthogonal.
This pattern is less emphasized for time windows near the start and the beginning of the audio snippets. We hypothesize that this is either because words are typically in the middle of each snippet and therefore relative position is more important there, or because the audio content at the start and end is less distinguishable.

\subsection{Latency Measurements}

We converted our KWT models, DS-CNN (with stride) \cite{zhang2017hello}, TC-ResNet \cite{choi2019temporal} and MHAtt-RNN \cite{rybakov2020streaming} to Tensorflow (TF) Lite format to measure inference latency on a OnePlus 6 mobile device based on the Snapdragon 845 (4x Arm Cortex-A75, 4x Arm Cortex-A55) and report accuracy figures for the Google Speech Commands V2 with 12 labels and 35 labels \cite{speechv2,rybakov2020streaming}. The TFLite Benchmark tool \cite{TFLite_benchmark} is used to measure latency, defined by the processing time of a single one-second input. For each model, we do 10 warmup runs followed by 100 inference runs, capturing the average latency on a single thread.

\begin{figure}[t]
    \centering
    \includegraphics[width=\linewidth]{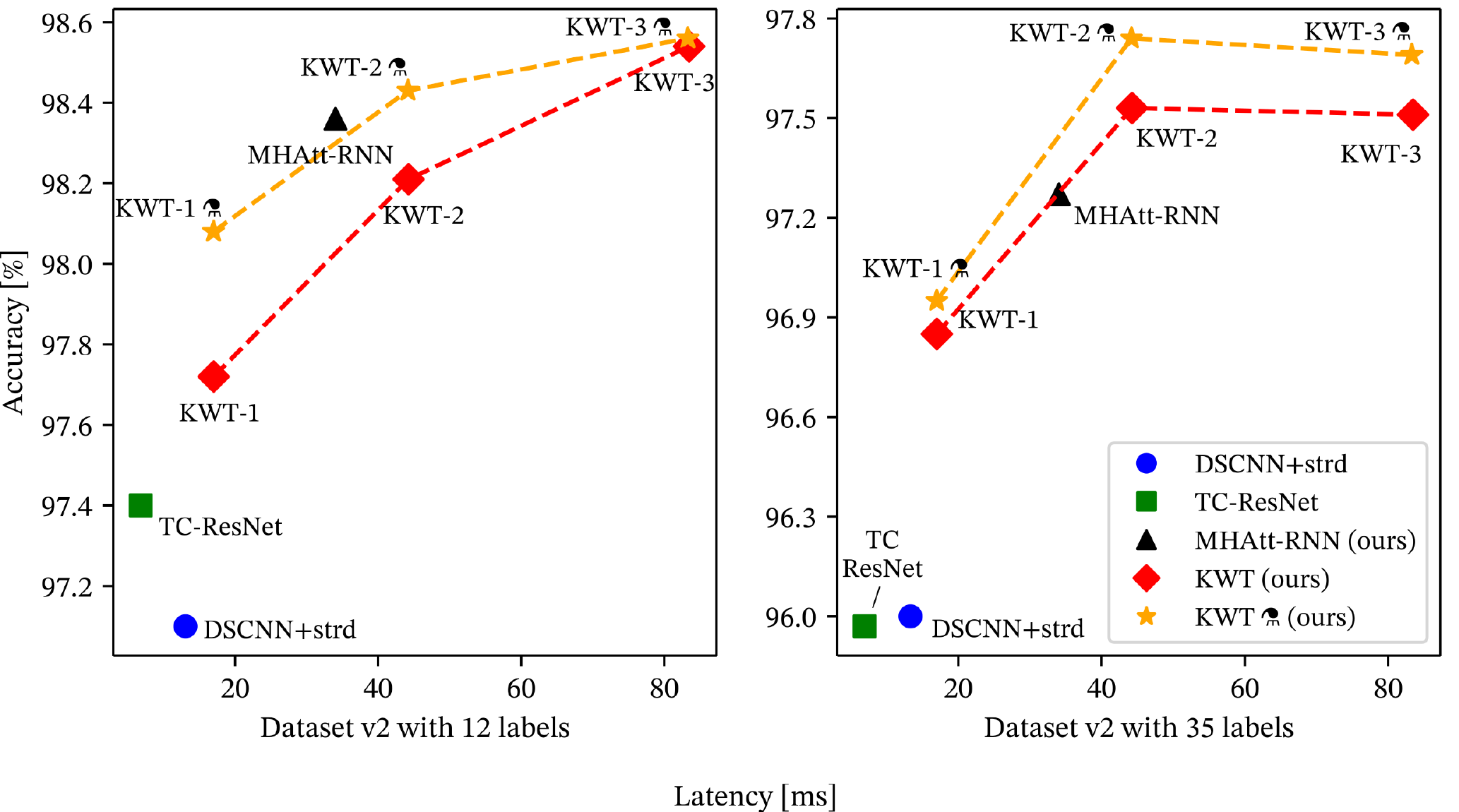}
    \caption{Latency and accuracy of processing a one-second input, on a single thread on a mobile phone.}
    \label{fig:latency}
  \end{figure}

In Figure \ref{fig:latency} we observe that Transformer-based models are competitive with the existing state-of-the-art despite being designed with no regard to latency. There is a broad body of research on optimizing Transformer models --- of particular note is the replacement of layer normalization and activations in \cite{sun2020mobilebert} that decreases latency by a factor of three. Our findings here suggest  many of these results could be leveraged in the keyword spotting domain to extend the practicality of these models.

\section{Conclusion}
In this paper we explore the direct application of Transformers to keyword spotting, using a standard architecture and a principled approach to converting the audio input into tokens.

In doing so we introduce KWT, a fully-attentional model that matches or exceeds the state-of-the-art over a range of keyword spotting tasks with real-world latency that remains competitive with previous work.

These surprising results suggest that Transformer research in other domains offers a rich avenue for future exploration in this space. In particular we note that Transformers benefit from large-scale pre-training \cite{dosovitskiy2020image}, have seen 5.5x latency reduction through model compression \cite{sun2020mobilebert} and up to 4059x energy reduction through sparsity and hardware codesign \cite{wang2021spatten}. Such improvements would make a meaningful impact on keyword spotting applications and we encourage future research in this area.

\section{Acknowledgements}
This work was partially supported by the Wallenberg AI, Autonomous Systems and Software Program (WASP), funded by the Knut and Alice Wallenberg Foundation. We thank Matt Mattina for supporting this work, Magnus Oskarsson for his feedback and comments, and Oleg Rybakov and Hugo Touvron for sharing their code with the community.

\bibliographystyle{IEEEtran}
\bibliography{mybib}

\end{document}